\documentclass[longauth]{aa}
\usepackage{natbib}
\usepackage{longtable,lscape}
\usepackage{amsmath}
\bibpunct{(}{)}{;}{a}{}{,}
\usepackage{graphicx}

\usepackage[colorlinks=true,citecolor=blue]{hyperref}
\usepackage{color}
\usepackage{xspace}
\usepackage{dcolumn}
\usepackage{longtable}
\usepackage{lscape}
\usepackage{wasysym}
\usepackage{rotating}
\usepackage{afterpage}
\usepackage{mathtools}

\newcommand{\MJup}{M$_{\mathrm{Jup}}$\xspace}

\newcommand{\MSun}{M$_{\odot}$\xspace}

\newcommand{\mic}{$\mu$m\xspace}
\newcommand{\as}{\hbox{$^{\prime\prime}$}\xspace}

\usepackage[varg]{txfonts}
\begin{document}

%
\title{Signs of late infall and possible planet formation around DR\,Tau using VLT/SPHERE and
    LBTI/LMIRCam.\thanks{Based on observation
    made with European Southern Observatory (ESO) telescopes at Paranal
    Observatory in Chile, under programs ID 0102.C-0453(A) and 1104.C-0416(A). It also make partial use of
    LBT/LMIRCam observations under program ID 74.}
}
\subtitle{}

\author{D. Mesa\inst{1}, C. Ginski\inst{2}, R. Gratton\inst{1}, S. Ertel\inst{3,4}, K. Wagner\inst{4,5}, 
    M. Bonavita\inst{1,6,7}, D. Fedele\inst{8,9}, M. Meyer\inst{10,11}, T. Henning\inst{12}, M. Langlois\inst{13,14}, A. Garufi\inst{8}, S. Antoniucci\inst{15}, R. Claudi\inst{1}, 
    D. Defr\`ere\inst{16}, S. Desidera\inst{1}, M. Janson\inst{17}, N. Pawellek\inst{18,19}, E. Rigliaco\inst{1}, V. Squicciarini\inst{20,1}, A. Zurlo\inst{13,22,23}, A. Boccaletti\inst{24}, M. Bonnefoy\inst{25}, F. Cantalloube\inst{13}, 
    G. Chauvin\inst{25}, M. Feldt\inst{12}, J. Hagelberg\inst{26}, E. Hugot\inst{13}, A.-M. Lagrange\inst{25}, 
    C. Lazzoni\inst{1,20}, D. Maurel\inst{25}, C. Perrot\inst{24,27,28}, C. Petit\inst{29}, D. Rouan\inst{24}, A. Vigan\inst{13} 
    }

\institute{\inst{1}INAF Osservatorio Astronomico di Padova, Vicolo dell'Osservatorio 5, Padova, Italy, 35122-I \\
    \inst{2}Anton Pannekoek Institute for Astronomy, University of Amsterdam, Science Park 904, 1098 XH Amsterdam, The Netherlands \\
    \inst{3}Large Binocular Telescope Observatory, 933 North Cherry Avenue, Tucson, AZ 85721, US \\
    \inst{4}Steward Observatory, Department of Astronomy, University of Arizona, 993 N. Cherry Ave, Tucson, AZ, 85721, USA \\
    \inst{5}NASA Hubble Fellowship Program - Sagan Fellow \\
    \inst{6}Institute for Astronomy, University of Edinburgh, EH9 3HJ, Edinburgh, UK \\
    \inst{7}Scottish Universities Physics Alliance (SUPA), Institute for Astronomy, University of Edinburgh, Blackford Hill, Edinburgh EH9 3HJ, UK \\
    \inst{8}INAF, Osservatorio Astrofisico di Arcetri, Largo Enrico Fermi 5, 50125, Firenze, Italy \\
    \inst{9}INAF, Osservatorio Astrofisico di Torino, Via Osservatorio 20, I-10025, Pino Torinese, Italy \\
    \inst{10}Astronomy Department, University of Michigan, Ann Arbor, MI 48109, USA \\
    \inst{11}Institute for Particle Physics and Astrophysics, ETH Zurich, Wolfgang-Pauli-Strasse 27, 8093 Zurich, Switzerland \\
    \inst{12}Max Planck Institut für Astronomie, Königstuhl 17, 69117 Heidelberg, Germany \\
    \inst{13}Aix Marseille Universit\'e, CNRS, LAM (Laboratoire d’Astrophysique de Marseille) UMR 7326, 13388 Marseille, France \\
    \inst{14} CRAL, UMR 5574, CNRS, Universit\'e de Lyon, Ecole Normale Sup\,erieure de Lyon, 46 All\'ee d’Italie, 
    F–69364 Lyon Cedex 07, France \\
    \inst{15}INAF-Osservatorio Astronomico di Roma, Via di Frascati 33, 00078 Monte Porzio Catone, Italy \\
    \inst{16}Institute of Astronomy, KU Leuven, Celestijnlaan 200D, 3001 Leuven, Belgium \\
    \inst{17}Department of Astronomy, Stockholm University, AlbaNova University Center, SE-109 91 Stockholm, Sweden \\
    \inst{18}Konkoly Observatory, Research Centre for Astronomy and Earth Sciences, Konkoly-Thege Miklós út 15-17, H-1121 Budapest, Hungary \\
    \inst{19}Institute of Astronomy, University of Cambridge, Madingley Road, Cambridge CB3 0HA, UK \\
    \inst{20}Department of Physics and Astronomy "Galileo Galilei", University of Padova, Italy \\
    \inst{22}N\'ucleo de Astronom\'ia, Facultad de Ingenier\'ia y Ciencias, Universidad Diego Portales, Av. Ejercito 441, Santiago, 
    Chile \\
    \inst{23}Escuela de Ingenier\'ia Industrial, Facultad de Ingenier\'ia y Ciencias, Universidad Diego Portales, Av. Ejercito 441, Santiago, Chile \\
    \inst{24}LESIA, Observatoire de Paris, Universit\'e PSL, CNRS, SorbonneUniversit\'e, Univ. Paris Diderot, Sorbonne Paris Cit\'e, 5 Place Jules
    Janssen, 92195 Meudon, France \\
    \inst{25}Univ. Grenoble Alpes, CNRS, IPAG, F-38000 Grenoble, France \\
    \inst{26}Geneva Observatory, University of Geneva, Chemin des Mailettes 51, 1290 Versoix, Switzerland \\
    \inst{27}Instituto de Física y Astronomía, Facultad de Ciencias, Universidad de Valparaiso, Av. Gran Bretaña 1111, Valparaíso, Chile \\
    \inst{28}N\'ucleo Milenio Formaci\'on Planetaria – NPF, Universidad de Valparaiso, Av. Gran Bretaña 1111, Valparaíso, Chile \\
    \inst{29}DOTA, ONERA, Universit\'e Paris Saclay, F-91123, Palaiseau France
}

   \date{Received  / accepted }

\abstract
{Protoplanetary disks around young stars often contain substructures like rings, gaps, and spirals that could be caused by interactions between the disk and forming planets. }
{We aim to study the young (1-3~Myr) star DR\,Tau in the near-infrared and characterize its disk, which was previously resolved through sub-millimeter interferometry with ALMA, and to search for possible sub-stellar companions embedded into it.}
{We observed DR\,Tau with VLT/SPHERE both in polarized light (H broad band) and total intensity (in Y, J, H, and K spectral bands). We also performed L' band observations with LBTI/LMIRCam 
on the Large Binocular Telescope (LBT). We applied differential imaging techniques to analyze the polarized data, using dual beam polarization imaging (DPI), and total intensity data, using both angular and spectral differential imaging (ADI, SDI).}
{We found two previously undetected spirals extending north-east and south of the star, respectively. We further detected an arc-like structure north of the star. Finally a bright, compact and elongated structure was detected at separation of $303\pm 10$~mas and position angle $21.2\pm 3.7$~degrees, just at the root of the north-east spiral arm. Since this feature is visible both in polarized light and in total intensity and has a flat spectrum it is likely caused by stellar light scattered by dust.}
{
The two spiral arms are at different separation from the star, have very different pitch angles, and are separated by an apparent discontinuity, suggesting they might have a different origin. The very open southern spiral arm might be caused by infalling material from late encounters with cloudlets into the formation environment of the star itself. The compact feature could be caused by interaction with a planet in formation still embedded in its dust envelope and it could be responsible for launching the north-east spiral. We estimate a mass of the putative embedded object of the order of few \MJup.}
   
   \keywords{Instrumentation: adaptive optics - Methods: data analysis - Techniques: imaging spectroscopy - Stars: planetary systems, Stars: individual: DR\,Tau}

\titlerunning{DR\,Tau}
\authorrunning{Mesa et al.}
   \maketitle
%

\section{Introduction}
\label{intro}

Protoplanetary disks surrounding young stars are considered the formation environment for planets \citep[see e.g., ][]{2012ApJ...756..133C,2014A&A...565A..15M}. Observing planets in formation has, however, so far been possible only for one 
confirmed case. This is that of PDS\,70 where two giant, forming planets have been imaged \citep[e.g., ][]{2018A&A...617A..44K,2018A&A...617L...2M,2018ApJ...863L...8W,2019NatAs...3..749H,2019A&A...632A..25M}. Other possible cases like e.g., that of HD\,169142 \citep[e.g., ][]{2017ApJ...850...52P,2018MNRAS.473.1774L,2019A&A...623A.140G} still need confirmation. Observations in the near-infrared (NIR) with instruments like SPHERE at VLT \citep{2019A&A...631A.155B}, GPI at the Gemini Telescope \citep{2014PNAS..11112661M} and CHARIS at the Subaru Telescope \citep{2015SPIE.9605E..1CG} and at millimeter wavelengths with the Atacama Large Millimeter Array (ALMA) enabled resolution of a wealth of substructures in protoplanetary disks. Such substructures can be gaps and rings \citep[e.g., ][]{2016A&A...590L...7P,2017A&A...601A...7F,2018ApJ...869L..41A,2018A&A...610A..24F,2018ApJ...869L..49I}, cavities \cite[e.g., ][]{2017AJ....154...33A,2017A&A...607A..55V,2021MNRAS.502.5779N} and spirals \citep[e.g., ][]{2012ApJ...748L..22M,2017A&A...601A.134M,2020A&A...637L...5B,2021ApJ...908L..25G,2021arXiv210713560B}. One of the most common explanations for these structures is the interaction between the disk and an unseen companion embedded in the disk itself \citep[see e.g., ][]{2016ApJ...818...76J,2018MNRAS.473.4459F,2018A&A...612A.104F}. However, alternative models have been proposed to explain these structures such as possible accumulation of dust at the snow lines \citep[e.g., ][]{2015ApJ...806L...7Z}, zonal flows \citep{2017A&A...600A..75B}, secular gravitational instability \citep{2014ApJ...794...55T}, magneto-rotational instability in the outer region of the disk \citep{2015A&A...574A..68F} or late infall of material on the disk \citep{2019A&A...628A..20D,2020A&A...633A...3K}. \par
For what concerns the spiral patterns in protoplanetary disks, hydrodynamical simulations \citep[see e.g., ][]{2015MNRAS.453.1768P} give strong indications that disk-planet interactions, in the early phases of the planetary formation, can produce both an inner and an outer spiral pattern through the presence of Lindblad resonances \citep[e.g., ][]{2013ApJ...779...59G}. Yet we still lack clear observational evidence to confirm these theories even if recent observations hint in that direction. Indeed \citet{2020A&A...636L...4M} were able to precisely define the position and the mass of a possible planet responsible for launching the spiral pattern in the disk of SR\,21. Moreover, \citet{2020A&A...637L...5B} proposed that the spirals in the AB\,Aur disk could be due to the presence of two low mass companions still in formation and identified two features that could correspond to those companions. Finally, the formation of
a spiral pattern driven by a stellar companion was proposed for the  binary star HD\,100453\,AB \citep{2016ApJ...816L..12D,2018ApJ...854..130W} \par
In this work we present new observations both in polarized light and total intensity of the system of DR\,Tau obtained with VLT/SPHERE and thermal infrared observations obtained with LBTI/LMIRCam. We found that DR~Tau is a very interesting system in this context because of its young age, the properties of its disk, and of its favourable almost pole-on orientation. The paper is organized as follows: in Section~\ref{target} we summarise the results of previous studies regarding the system of DR\,Tau while in Section~\ref{s:obs} we present our new observations of the system and the data reduction methods adopted. In Section~\ref{s:res} we present our results that are then discussed in Section~\ref{s:dis}. Finally, in Section~\ref{s:conclusion} we give our conclusions.


\section{The target}
\label{target}


DR\,Tau is a very active classical T\,Tauri star (CTTS) situated in the Taurus-Auriga star forming region but its estimated distance of 192.97$\pm$1.23~pc \citep{2021A&A...649A...1G} is larger than that of the main region ($\sim$140~pc). It has shown a slow increase of its brightness between 1960 and 1980 \citep{1979A&A....79L..18C,1980IBVS.1747....1G} passing from being a faint star with a V band magnitude of 14~mag to one of the brightest stars of the association (V$\sim$11~mag). This was probably caused by a strong accretion process \citep{1988ApJ...330..350B}. Due to this, DR\,Tau has also been classified as an EXOr variable \citep[see e.g., ][]{1989ESOC...33..233H,2009ApJ...693.1056L}. Since then it has maintained its augmented flux but displaying large photometric and spectral variability \citep{2001AJ....122.3335A,2007A&A...461..183G}.

Its spectral classification is quite uncertain due to the high and variable veiling produced by the accretion process. It varies between K5V and M0V with a larger number of earlier spectral classification \citep{2011A&A...535A...6P, 2014ApJ...780...26B,2019ApJ...882...49L} with respect to the later ones \citep[e.g. ][]{2019A&A...632A..32M}. Also its mass suffers from similar uncertainties ranging from 0.4 \citep[e.g. ][]{2019ApJ...874...24S} to more than 1~\MSun \citep[e.g. ][]{2013ApJ...771..129A}. Recently, exploiting the measurement of the Keplerian rotation of the disk \citet{2021ApJ...908...46B} found a mass of $1.18^{+0.59}_{-0.44}$~\MSun for DR\,Tau. In the same work, however, they also found slightly subsolar masses for the star using various sets of evolutionary models. The age estimates of the system vary between 0.9~Myr obtained by \citet{2019A&A...632A..32M} using the \citet{2000A&A...358..593S} evolutionary tracks and 3.2~Myr obtained by \citet{2019ApJ...882...49L} adopting the pre-main-sequence evolutionary models by \citet{2015A&A...577A..42B} and \citet{2016A&A...593A..99F}.

The emission from its disk has been widely studied resulting in the first discovery of H$_2$O spectro-astrometric signatures in a protoplanetary disk \citep{Brown2013}. At the same time, other important molecular lines were detected in emission, such as CO lines \citep{2019ApJ...882...49L}. For these molecules single peaked lines were found in contrast with what is expected for a Keplerian disk. A possible explanation could be a low disk inclination coupled with the presence of a slow disk wind \citep{2011A&A...527A.119B,2011ApJ...733...84P, 2019ApJ...874...24S}. The dust disk was recently resolved at a wavelength of 1.3~mm using ALMA by \citet{2019ApJ...882...49L} finding a total disk flux density of $127.18^{+0.20}_{-0.22}$~mJy, a radius of 0.267\as ($\sim$51~au), an inclination of $5.4^{\circ}$$^{+2.1}_{-2.6}$, and a position angle of $3.4^{\circ}$$^{+8.2}_{-8.0}$. In contrast, a much larger radius of 246~au was found for the gas by \citet{2021ApJ...908...46B}.


\section{Observations and data reduction}
\label{s:obs}

\begin{table*}[!htp]
  \caption{List and main characteristics of the observations of DR\,Tau used for this work.}\label{t:obs}
\centering
\begin{tabular}{ccccccccc}
\hline\hline
Date  &  Obs. mode & Coronograph & DIMM seeing & $\tau_0$ & wind speed & Field rotation & DIT & Total exp\\
\hline
2018-11-26  & IRDIS DPI    & N\_ALC\_YJH\_S & 0.28\as & 8.0 ms & 2.98 m/s & 0              &  32 s    &   640 s \\
2019-11-28  & IRDIFS\_EXT  & N\_ALC\_Ks     & 0.51\as & 5.3 ms & 2.75 m/s &$26.16^{\circ}$ &  96 s    &  4608 s \\
2020-02-02  & LBT/LMIRCAM  & AGPM           & 1.10\as & ///    & 9.70 m/s &$75.50^{\circ}$ & 1.01589s &  2660 s \\
\hline
\end{tabular}
\end{table*}

DR\,Tau was observed with SPHERE \citep{2019A&A...631A.155B} at the ESO Very Large Telescope (VLT) in two epochs and with the L- and M-band Infrared Camera 
\citep[LMIRCam; ][]{2010SPIE.7735E..3HS,2016SPIE.9907E..04H} of the Large Binocular Telescope Interferometer (LBTI) in one epoch. These observations are described below and detailed in Table~\ref{t:obs}.

\subsection{SPHERE/VLT polarized imaging}
\label{s:polar}

\begin{figure}
\centering
\includegraphics[width=1.\columnwidth]{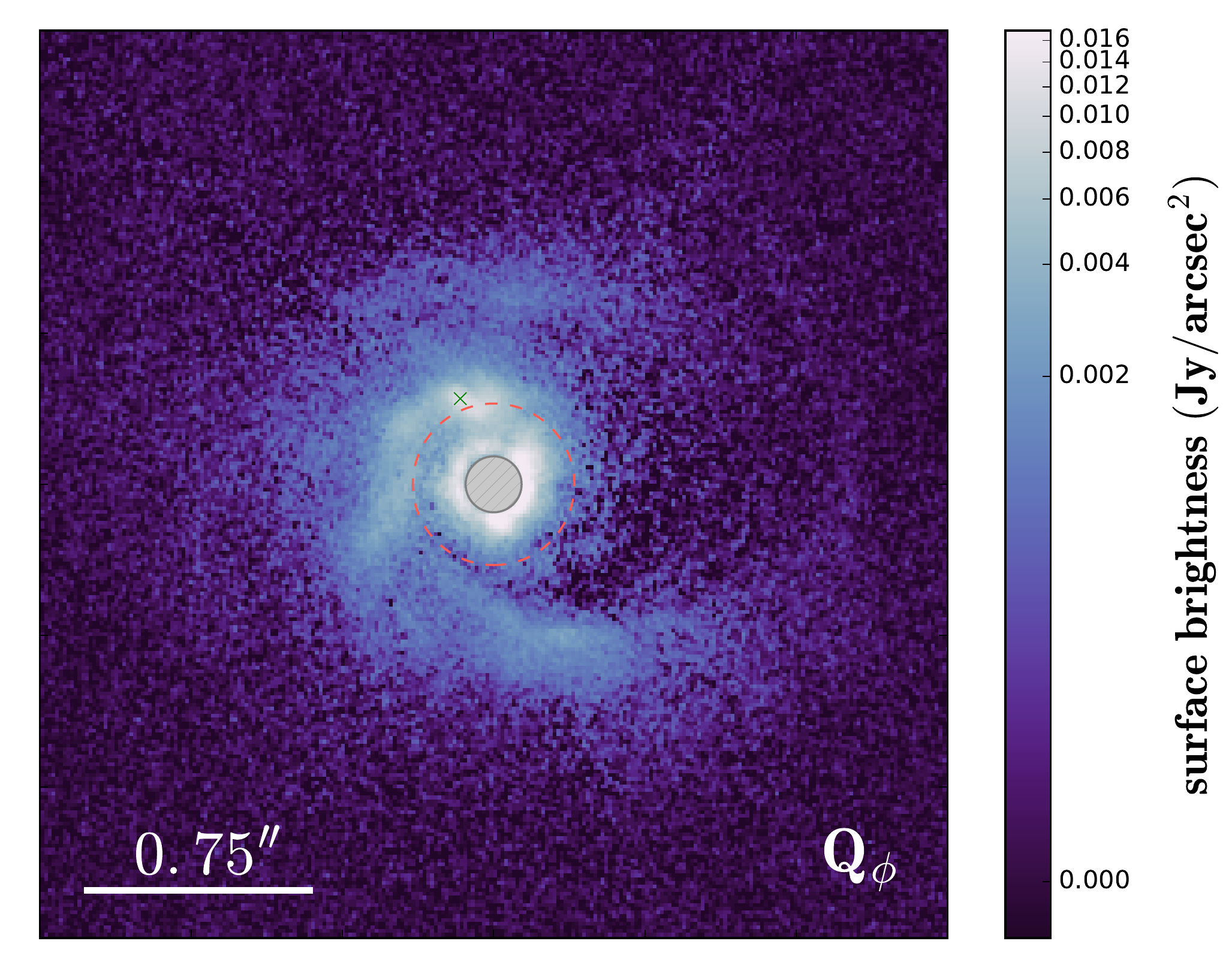}
\caption{SPHERE H-band $Q_{\phi}$ final image obtained from the DPI data. The red dashed circle represents the radius found from ALMA data. The magenta cross indicates the position of the point source candidate described in this paper (see Section~\ref{s:morpho}).
The grey hashed circle in the image centre marks the size of the coronagraph. No deprojection to the disk plane was applied to this image.}
\label{f:qphi}
\end{figure}

In the first epoch, during the night of 2018-11-26,  SPHERE was operating in dual beam polarization imaging \citep[DPI; ][]{2020A&A...633A..63D,2020A&A...633A..64V} using the IRDIS \citep{2008SPIE.7014E..3LD} infrared camera in the broad-band H filter ($\lambda$=1.625~\mic, $\Delta\lambda$=0.29~\mic). The observation was performed in field stabilized mode. 
We obtained five polarimetric cycles to measure the linear polarimetric parameters $Q^{+}$, $Q^{-}$, $U^{+}$ and $U^{-}$. The weather conditions were excellent during all the observation as detailed in Table~\ref{t:obs}.

The data were reduced using the IRDAP \citep[IRDIS data reduction for accurate polarimetry; ][]{2017SPIE10400E..15V,2020A&A...633A..64V} pipeline. As a first step the pipeline pre-processes raw data applying dark subtraction, flat fielding, bad pixel correction and registering each image. In a second step we derive the linear Stokes parameters Q and U and the relative total intensities that are then corrected for instrumental polarization effects and for cross talk. Finally, the pipeline computes the linearly polarized intensity, angle, and  degree of polarization for the source. From these values it provides the azimuthal Stokes parameters $Q_{\phi}$ and $U_{\phi}$ \citep[see ][ for a definition of $Q_{\phi}$ and $U_{\phi}$]{2020A&A...633A..63D} that in this context represent the polarimetric signal and the relative noise, respectively.

The final $Q_{\phi}$ image is displayed in Figure~\ref{f:qphi} where we also plot the dust radius found from ALMA data (red dashed circle) and the position of the compact feature identified in this work which will be described in Section~\ref{s:morpho} (magenta cross). The location of the SPHERE coronagraph is also shown with a grey disk.

\subsection{SPHERE/VLT total intensity imaging}
\label{s:sphere}

\begin{figure*}[htp]
\centering
\includegraphics[width=1.\textwidth]{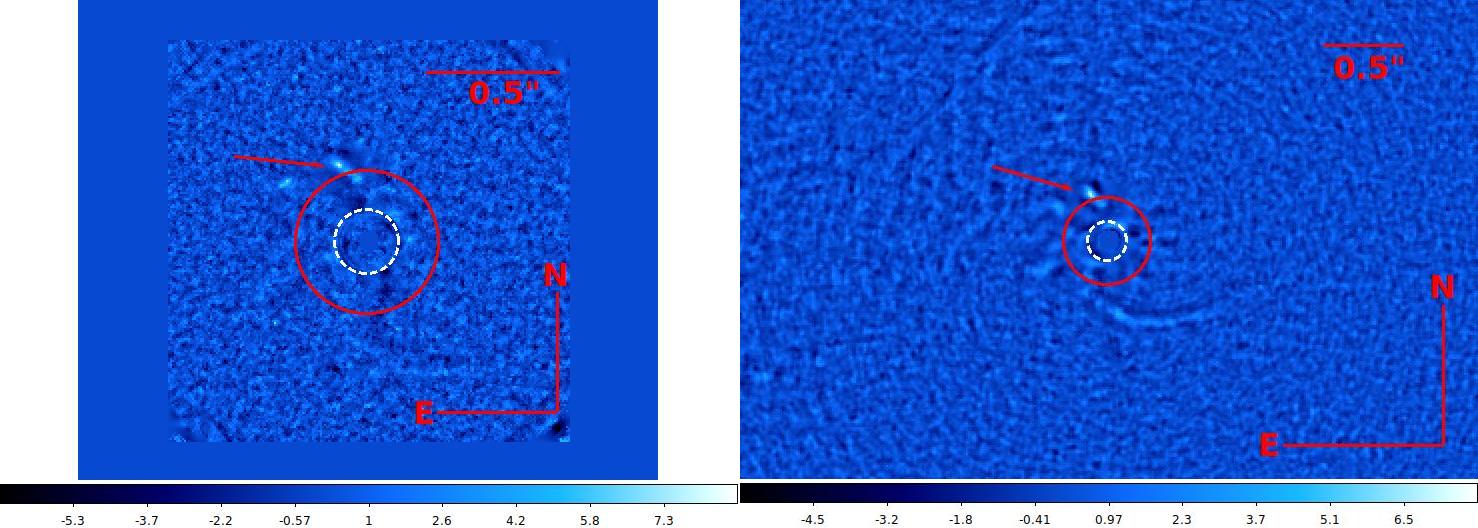}
\caption{{\it Left:} Final signal-to-noise map obtained from IFS data obtained using the ASDI-PCA subtracting 10 principal components. {\it Right:} Final  signal-to-noise map obtained from IRDIS data obtained using the PCA subtracting 2 principal components. In both images the red circle represents the disk radius of the inner disk as found with ALMA data and the red arrow indicates the position of the compact structure described in Section~\ref{s:morpho}. Moreover, the white dashed circle represents
the inner working angle ($\sim$0.09\as) of the coronagraph used for these observations. The colour bars below the two images indicates the
S/N values obtained from our data reduction procedure. As for the case of Figure~\ref{f:qphi}, no deprojection to the
real orientation of the disk plane was applied to both images.}
\label{f:nopol}
\end{figure*}

The second epoch observation of DR\,Tau with SPHERE was performed on the night of 2019-11-28 in the context of the SHINE \citep[SpHere INfrared survey for Exoplanets; ][]{2017sf2a.conf..331C,2021A&A...651A..70D,2021A&A...651A..71L,2021A&A...651A..72V} survey. The observation was performed using the IRDIFS\_EXT observing mode with the integral field spectrograph \citep[IFS; ][]{2008SPIE.7014E..3EC} operating in Y, J and H spectral bands (between 0.95 and 1.65~\mic) and with IRDIS using the K band with the K12 filter pair \citep[wavelength K1=2.110~\mic; wavelength K2=2.251~\mic; ][]{2010MNRAS.407...71V}. \par
We also obtained frames with satellite spots symmetric with respect to the central star before and after the coronagraphic sequences. This enabled us to determine the position of the star behind the coronagraphic focal plane mask and accurately recentre the data \citep{2013aoel.confE..63L}. Furthermore, to be able to correctly calibrate the flux of companions, we acquired images with the star off-axis. In these cases, an appropriate neutral density filter was used to avoid saturation of the detector.

The data were reduced through the SPHERE data centre \citep{2017sf2a.conf..347D} applying the appropriate calibrations following the data reduction and handling \citep[DRH; ][]{2008SPIE.7019E..39P} pipeline. In the IRDIS case, the calibrations included the dark and flat-field correction and the definition of the star centre. In addition to the dark and flat-field corrections, IFS calibrations included the definition of the position of each spectra on the detector, the wavelength calibration, and the application of the instrumental flat field. On the pre-reduced data we then applied speckle-subtraction algorithms like TLOCI \citep{2014SPIE.9148E..0UM} and principal components analysis \citep[PCA; ][]{2012ApJ...755L..28S} as implemented in the consortium pipeline application SpeCal \citep[Spectral Calibration; ][]{2018A&A...615A..92G}. For IFS data, we used the Angular-Spectral Differential Imaging (ASDI) PCA approach that uses the 4-d datacube (spatial dimension, time, and wavelength) as described in \citet{2014A&A...572A..85Z} and in \citet{2015A&A...576A.121M}.

The final signal-to-noise ratio (SNR) maps obtained from this procedure are shown in Figure~\ref{f:nopol} both for IFS (left panel) and IRDIS (right panel). Like for the polarimetric data case, we overplot on both images a red circle indicating the radius of the inner dust disk as found by ALMA data. The white dashed circle indicates in both images the extension of the SPHERE coronagraph for this observation.

\subsection{LMIRCam/LBT data}
\label{s:lmircam}

\begin{figure}
\centering
\includegraphics[width=1.\columnwidth]{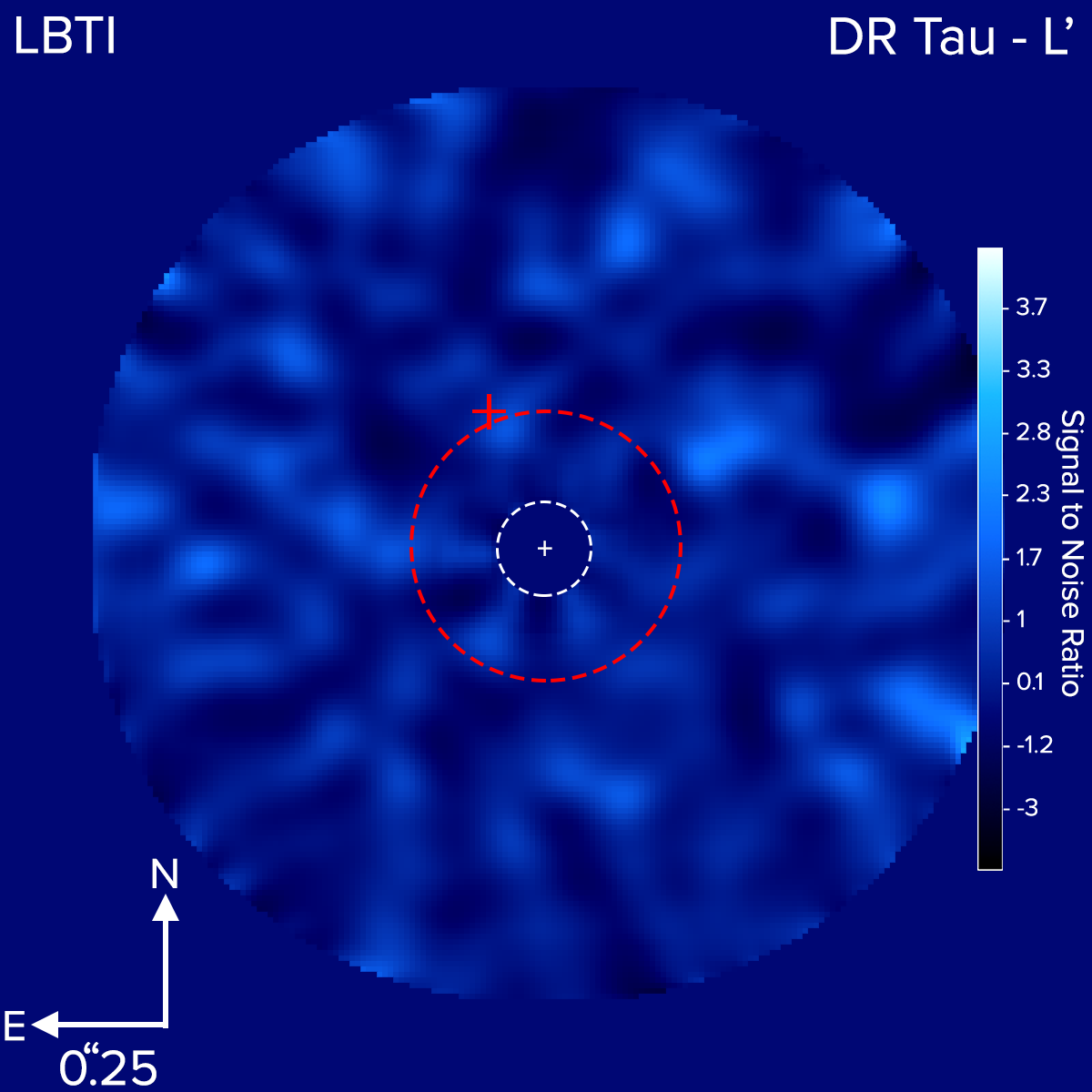}
\caption{Central part of the signal-to-noise map obtained from the LBTI/LMIRCam data using the KLIP algorithm and subtracting 10 
modes. The dashed red circle indicates the radius of the dust disk detected with ALMA while the dashed white circle represents the
inner working angle of the AGPM coronagraph (0.09\as). The position of the feature found with the SPHERE data and 
described in Section~\ref{s:res} is marked with a red cross. Like for the case of the SPHERE images this image has not been deprojected to the disk plane.}
\label{f:lbt}
\end{figure}

We also observed DR\,Tau in the L' band with LBTI/LMIRCam using the instrument's 
non-interferometric, individual-aperture, adaptive-optics imaging mode \citep{ 2020SPIE11446E..07E}. These observations were executed on the night of 2020-02-03 using only the left side of the LBTI and the Annular Groove Phase Mask (AGPM) coronagraph \citep{2014SPIE.9148E..3XD}. The observations were performed in pupil stabilized mode to allow to implement the angular differential imaging \cite[ADI; ][]{2006ApJ...641..556M}. The observations were designed to last $\sim$5~hours but high winds complicated the data acquisition and required terminating the observation after two hours.
The quality of the obtained data was affected by wind shake of the secondary mirror which rendered the adaptive optics loop unstable and prevented the use of the automated coronagraph centring 
loop \citep[QACITS approach, ][]{2015A&A...584A..74H} so that this step was performed manually at reduced effectiveness. \par
Basic LBTI calibrations include correlated double sampling, bad pixel correction, and sky/background subtraction. The frames were aligned with cross-correlation and centred using a rotational-based centring approach \citep{2015ApJ...815..108M} that utilizes the rotational symmetry of the PSF to determine the star’s location behind the coronagraph. Poor-quality frames (those with less than an 80\% maximum cross-correlation with the median pupil of the sequence) were then removed, resulting in $\sim$30\% frame rejection. The final dataset that was used for analysis includes 4525 frames covering 75.5$^\circ$ of field rotation. At this stage, synthetic planets were injected (when relevant) using the unsaturated and unocculted PSF of the star to assess the final image sensitivity. The spatial background was further mitigated by subtracting the mode of each column and applying a 15x15 pixel high-pass filter. The PSF was then modelled and subtracted using the KLIP algorithm \citep{2012ApJ...755L..28S} with 10 KL-modes over annular segments of 60$^\circ$ azimuthal range and a radial range of 10-100 pixels from the image centre ($\sim$0.1\as-1.1\as). Finally, the frames were derotated and combined using a noise-weighting approach \citep{2017RNAAS...1...30B}. The combined image obtained from this procedure is displayed in Figure~\ref{f:lbt}, in which we also show the location of the dust disk radius observed by ALMA with a red dashed circle and the position of the compact feature detected using SPHERE and described in Section~\ref{s:morpho} with a red cross. The white dashed circle indicate the location of the AGPM coronagraph used for this observation.
No source is detected with SNR$\geq$3.




\section{Results}
\label{s:res}

\subsection{Disk morphology}
\label{s:morpho}

\begin{figure*}[htp]
\centering
\includegraphics[width=1.\textwidth]{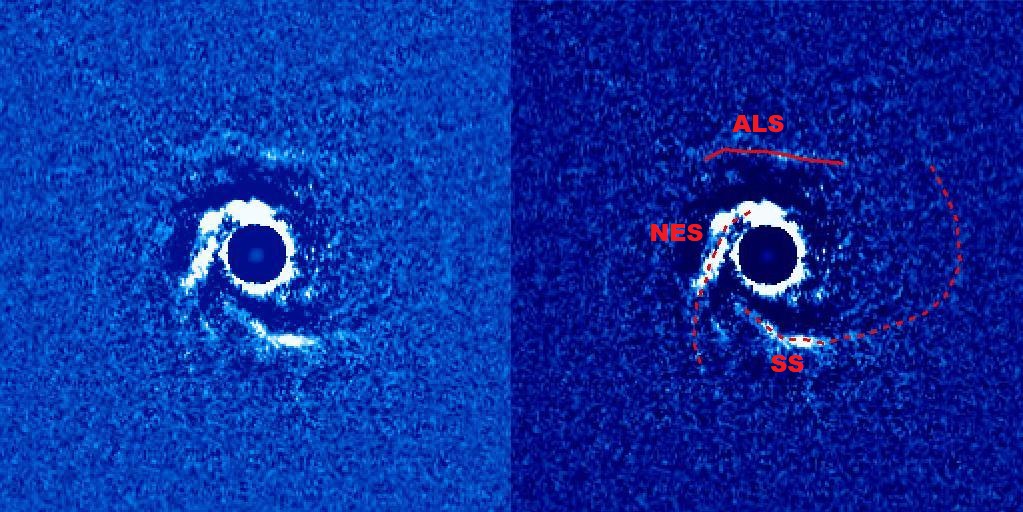}
\caption{{\it Left: }High pass filter of Figure~\ref{f:qphi} to highlight the presence of two spirals in the outer part of the disk. 
{\it Right: } Same as in the left panel of this Figure but with overplotted two red dashed lines to indicate the positions of the two 
spirals (labelled with NES for the north-eastern spiral and with SS for the southern spiral) and one red solid line to indicate the
position of the arc-like structure (labelled with ALS).}
\label{f:spiral}
\end{figure*}


\begin{figure}
\centering
\includegraphics[width=1.\columnwidth]{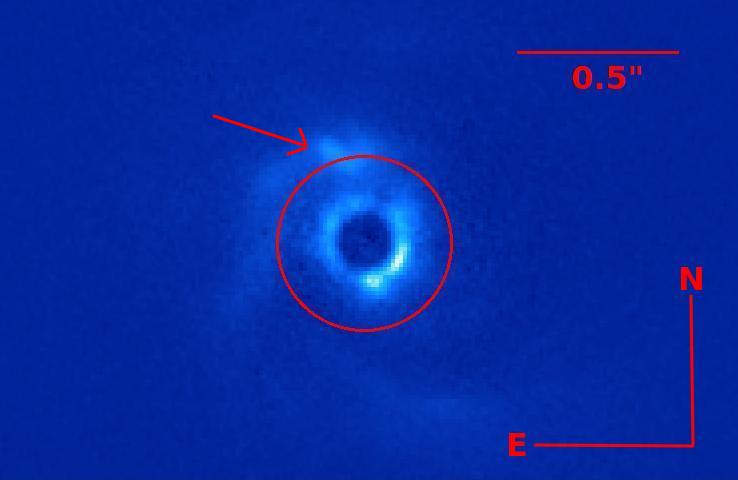}
\caption{Same $Q_{\phi}$ final image obtained from the DPI data as in Figure~\ref{f:qphi} but with different contrast settings and zoomed to make more evident the presence of the same feature visible in the high-contrast imaging data. The position of the feature is indicated by a red arrow.}
\label{f:qphizoom}
\end{figure}

\begin{figure}
\centering
\includegraphics[width=1.\columnwidth]{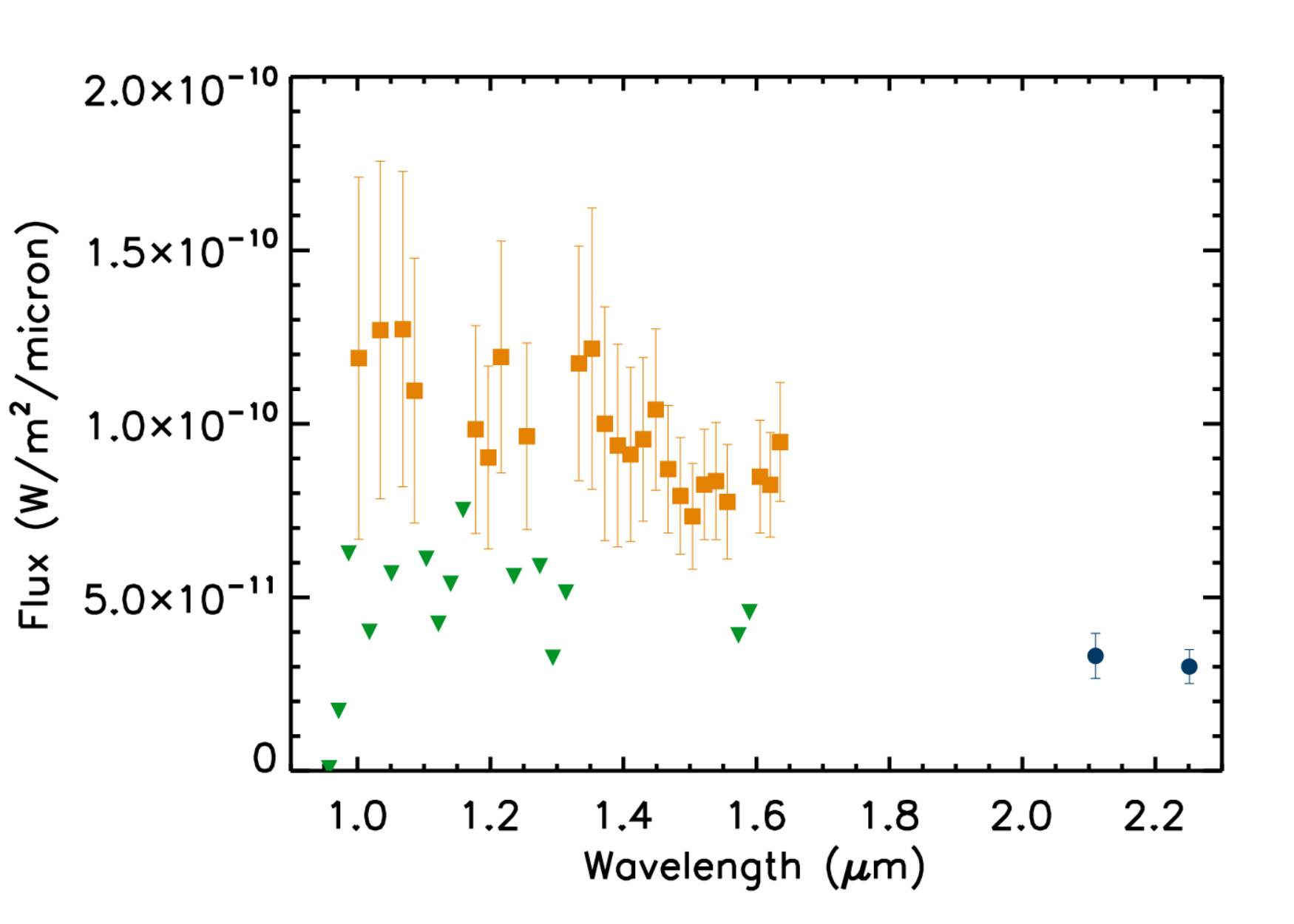}
\caption{Extracted spectrum of the elongated compact structure from the IFS and IRDIS data. The orange squares are the 
IFS data points, the blue circles are the IRDIS data points. The green upside-down triangles are the
upper limits obtained for some wavelengths.}
\label{f:ifsspectrum}
\end{figure}


\begin{figure*}[htp]
\centering
\includegraphics[width=0.49\textwidth]{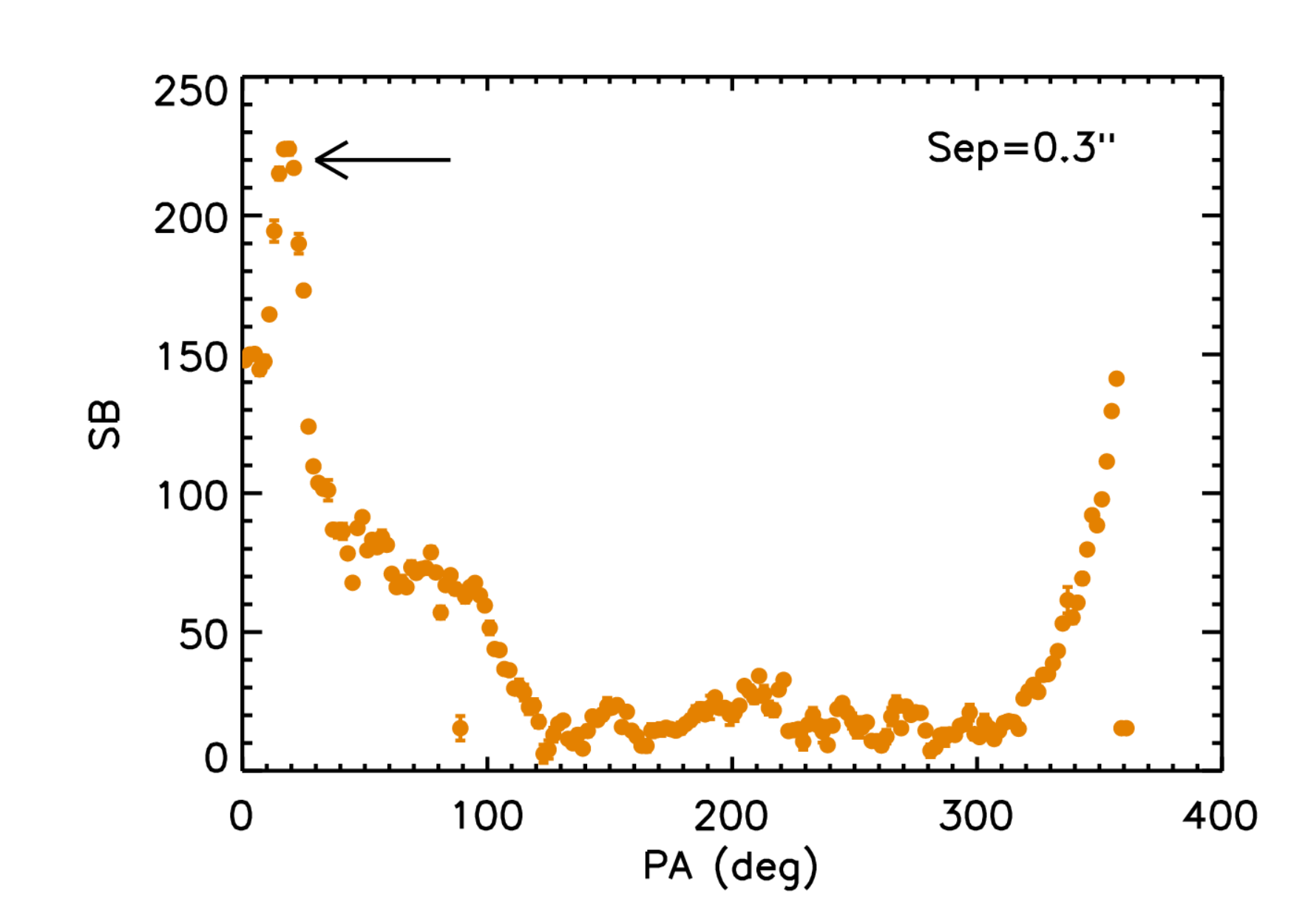}
\includegraphics[width=0.49\textwidth]{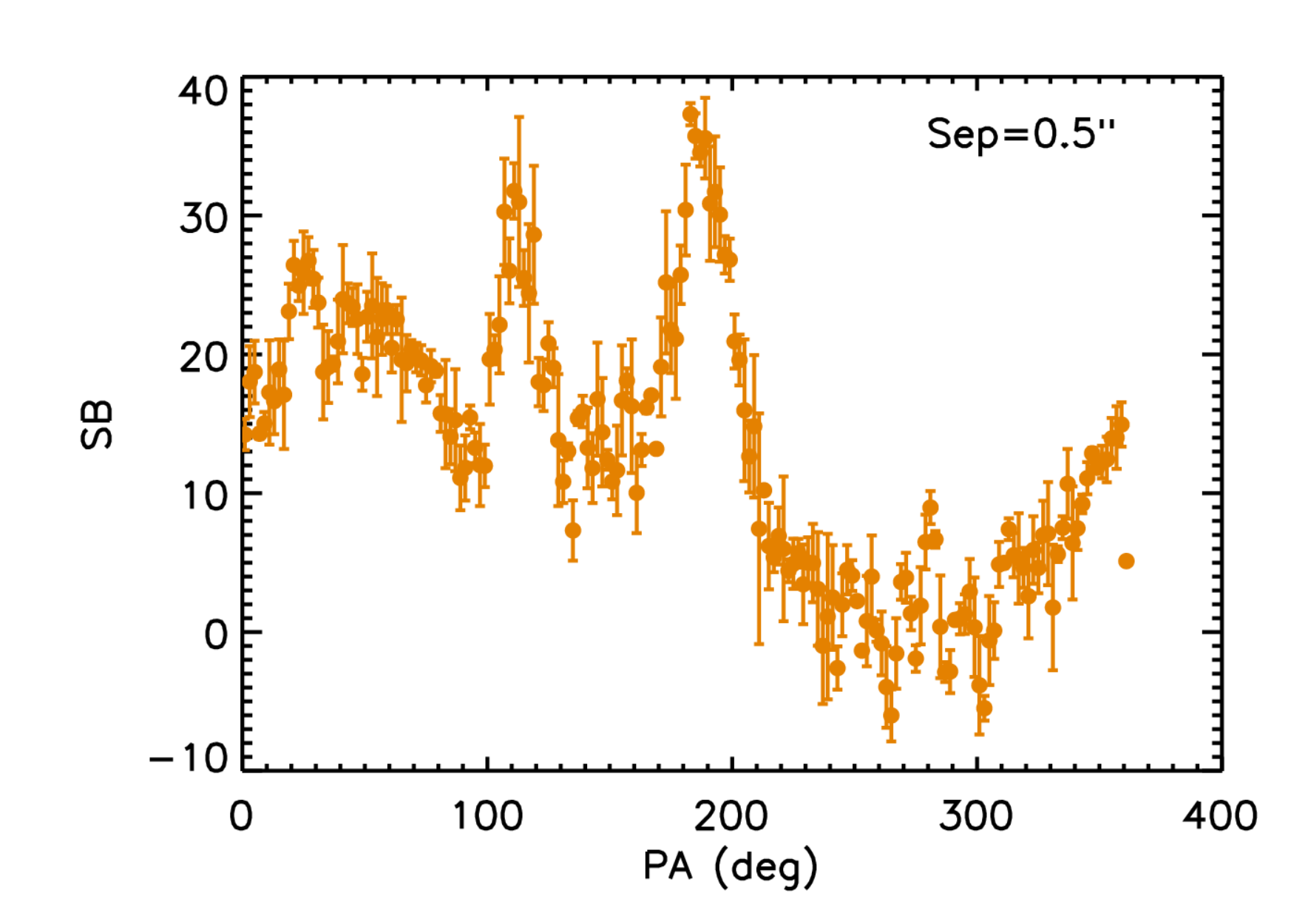}
\includegraphics[width=0.49\textwidth]{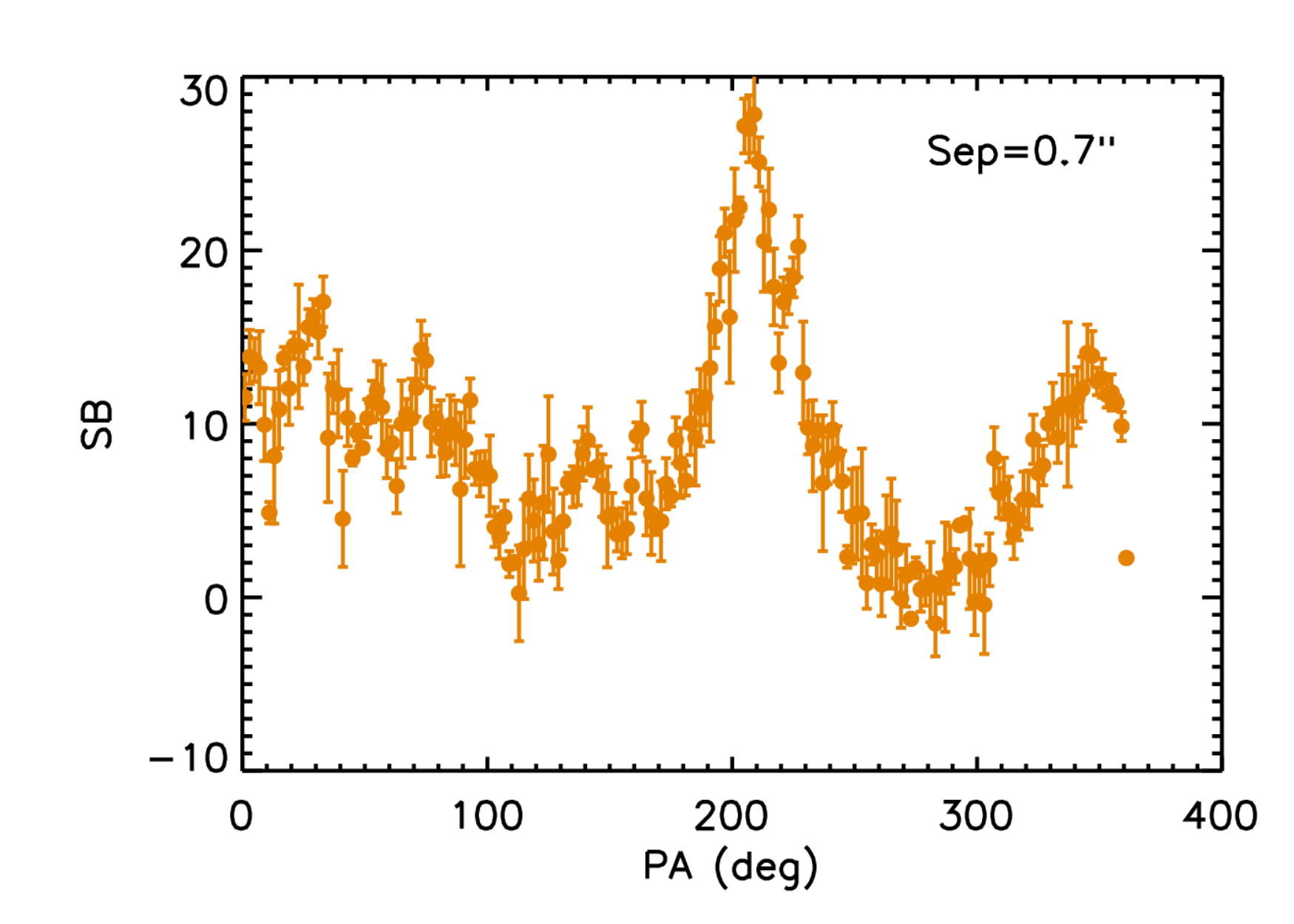}
\includegraphics[width=0.49\textwidth]{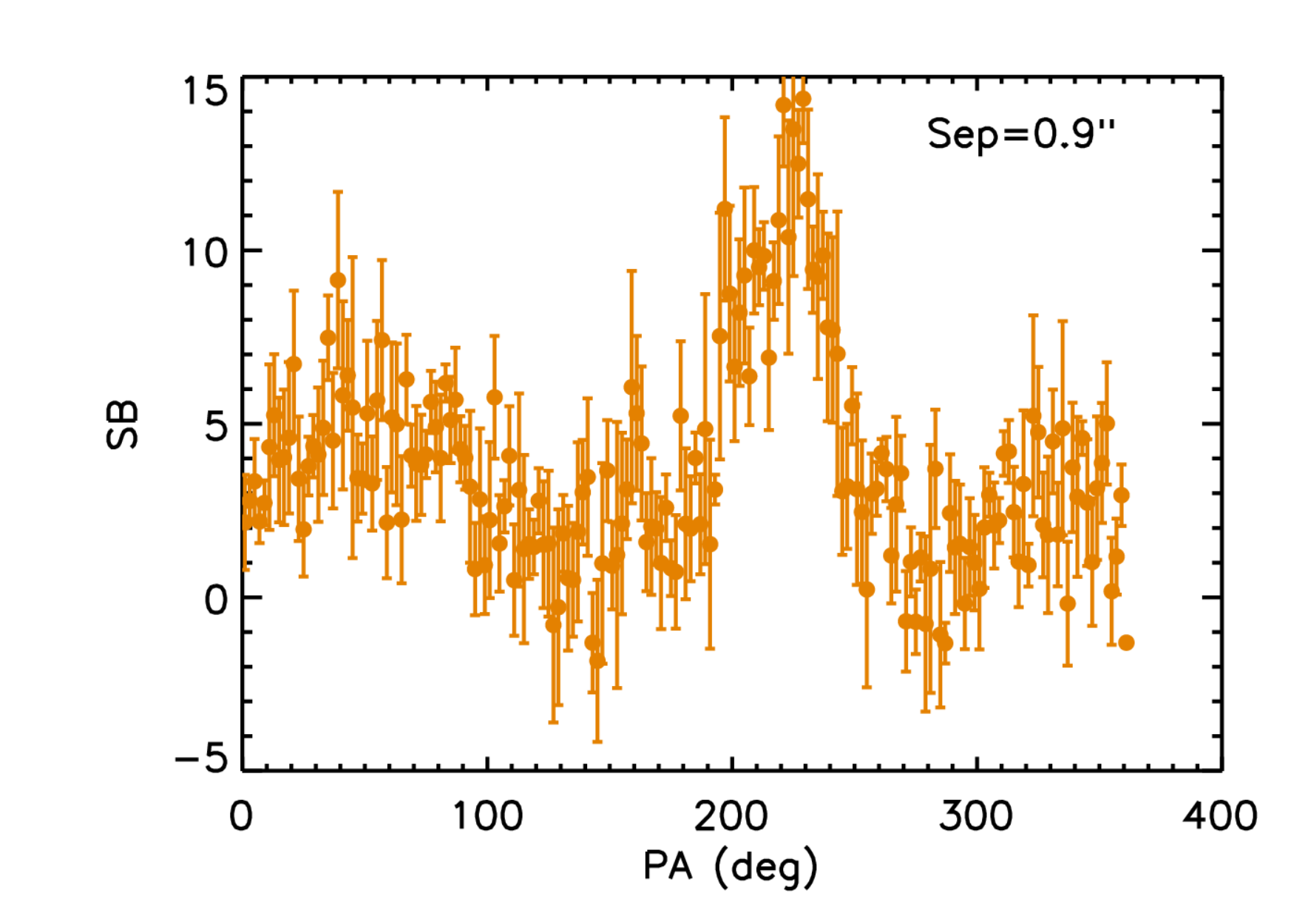}
\caption{Azimuthal surface brightness profiles obtained around DR\,Tau using the IRDIS polarized data at separations of 0.3\as (upper left panel), 0.5\as (upper right panel), 0.7\as (bottom left panel) and 0.9\as (bottom right panel). The black arrow in the top-left
panel indicates the position of the peak due to the compact feature. The error bars in the 0.3\as plot are barely visible due to the higher flux scale in that plot with respect to those obtained at larger separations.}
\label{f:aziprof}
\end{figure*}

The DR\,Tau circumstellar environment shown in Figure~\ref{f:qphi} and~\ref{f:nopol} is quite complex and can be subdivided in three most evident substructures: the inner disk, the spiral arms and a bright compact structure.

\subsubsection{Inner disk}

The polarized data (see Figure~\ref{f:qphi}) show that the inner region just around the star is populated by material (probably both dust grain and gas) at least at separation larger than the coronagraph inner working angle (that is 92.5~mas corresponding to $\sim$18~au at the distance of the system). To the west and to the south of the star the disk detected in scattered light has a radius comparable to that obtained from sub-mm emission with no or faint signal outside this separation. 

\subsubsection{Spiral arms}
\label{s:spiralarm}

We observe a strong emission north- and eastward of the star indicating the presence of a clump of material outside the radius of the 
disk detected by ALMA. As ALMA is sensitive to mm dust grains this emission could then be dominated by the presence of smaller grains.
We note however that the ALMA observations presented by \citet{2019ApJ...882...49L} are probably not deep enough to reveal 
extended dust emission at low surface brightness. New deeper ALMA observations will then be needed for a definitive conclusion on
this point.

A spiral pattern starts from the clump described above and at first look it seems to be composed by a single arm wrapped all around the star (see Figure~\ref{f:qphi}). However, a more careful analysis reveals that the system is likely 
composed of at least two different arms. 
In the left panel of Figure~\ref{f:spiral} we display the high-pass filtered version of Figure~\ref{f:qphi} to highlight the presence of these two arms. 
Their traces are shown for more clarity by two red dashed lines in the right panel of the same Figure. The first one of these arms starts from north-east of the star at a position angle (PA) of around $20^{\circ}$ and moves towards the south to end at the south-east of the star at a separation of $\sim$0.8\as (corresponding to a separation of 154~au at the distance of the system) 
and a PA of $\sim$$150^{\circ}$. The second spiral arm departs at a larger separation from the star than the first one ($\sim$0.35\as) and it is more wrapped. It starts 
just south from the star at a PA of $\sim$$150^{\circ}$ where it is very bright and it wraps around the star towards the west and then to the north fading away while it departs from the star. We are however able to trace it 
up to a separation of 1.14\as from the star and a position angle of $\sim$$300^{\circ}$, north-west of the star. This separation is $\sim$220~au at the distance of the system and it is 
consistent with the size of the gas disk observed by \citet{2021ApJ...908...46B}.

Additionally, an arc-like structure is visible north to the star at separations ranging between $\sim$0.63\as and $\sim$0.73\as and at PA ranging between around $-40^{\circ}$ and 
$\sim$$33^{\circ}$. Its position is highlighted by a red solid line in the right panel of Figure~\ref{f:spiral}.

The pitch angle of the two spirals can be determined transforming the polarized data image in polar coordinates. In this case the two spiral arms are seen as diagonal lines
and their inclination with respect to the vertical gives their pitch angle. For a nearly face-on disk such as that of DR\,Tau, this is expected to be rather constant along a 
spiral arm except in the close vicinity of an object possibly launching it, though spirals may well be generated by mechanisms other than the presence of a
companion. We found for the north-eastern spiral a moderate pitch angle of $\sim$$11^{\circ}$ while the southern spiral has a much larger starting pitch angle of $\sim$$26^{\circ}$ that makes it a very open spiral arm. Additionally, the pitch angle of the second spiral
is strongly variable with values increasing at larger separation and a clear bending toward the north at a separation of $\sim$1.15\as and at a PA of $\sim$$260^{\circ}$.


The two spirals are also visible both in IFS and IRDIS total intensity images from the second epoch even if with a very low SNR as evidenced in Figure~\ref{f:nopol}. In this case however the inner region is completely depleted of any signal. This might indicate that the material in the inner region around the star is strongly polarized. 

\subsubsection{Elongated compact feature}

The most striking structure in the total intensity images in Figure~\ref{f:nopol} is the elongated compact feature that is clearly visible both in IFS and in IRDIS images north to the star
(indicate by a red arrow in both images of Figure~\ref{f:nopol}). This feature is elongated especially in the case of the IFS image so that it is difficult to define its position with the usual methods considered for the direct imaging data. 
We however applied the negative planet method \citep{2011A&A...528L..15B,2014A&A...572A..85Z} as implemented in the SpeCal tool limiting its application only to the IRDIS case where the feature is less elongated. We note that the negative planet method assumes that the source is point-like. This is likely not true in this case but the method can anyhow give a good estimation
of the position of the compact structure. From this procedure we obtained a separation of 303$\pm$10~mas, corresponding to a projected separation of $\sim$58~au at the distance of the system, and a PA of $21.2^{\circ}\pm3.7^{\circ}$. The unusually high error bars on the SPHERE astrometry reflect the difficulties linked to the elongation of this feature described above.

The compact feature is not visible in Figure~\ref{f:qphi} due to the particular contrast settings chosen with the aim to fully display the extension of the spiral arms. Figure~\ref{f:qphizoom} shows the $Q_{\phi}$ final image adopting different contrast settings. Here, while the spiral arms are barely visible, the feature detected in the total intensity images is obvious and its position is highlighted by a red arrow. 
This feature is clearly associated with the region from which the north-eastern spiral is launched as highlighted by one of the red dashed lines in Figure~\ref{f:spiral}. 

The facts that this feature is quite elongated and is detected both in polarized and non-polarized light indicate that we are not 
probably directly observing the photosphere of a planet. This is further reinforced by the IFS and IRDIS spectrum extracted using 
the SpeCal tool and applying the negative planet method. The same limits of this method described above when defining the astrometry of this 
structure are valid also in this case but it is however useful to give indications about the shape of the spectrum. The results of this 
procedure is shown in Figure~\ref{f:ifsspectrum}. The signal-to-noise of this spectrum is very low and for a number of wavelengths it 
was only possible to obtain upper limits. However, the resulting spectrum appears blue and this is a further indication that we 
are looking at stellar light reflected by dust. 

In principle, having two different epochs separated by nearly one year from each other we could determine the motion of this compact feature. In any case the astrometric measures are very complicated due to the elongated shape of the structure as explained above. As a consequence the error bars are very large so that different measures are within error bars each other. Also, the extraction of the astrometric measure from the polarimetric data is further hampered by the noisy environment because of the presence of polarized
material around the structure itself. Moreover, we have to consider that the two observations were not done with the same instrumental configuration. In particular the two IRDIS observations were done at two different spectral bands (H broad-band and K1K2 dual imaging for the polarized and non-polarized observations). The fact that we are observing emission from a dusty environment implies that we are probably observing at a different optical thickness when using different wavelengths. All these considerations make the astrometric comparison of measurements taken at different wavelengths very challenging. 
We then decided to limit ourselves to the detection of the rotation of this feature comparing the polarimetric data with those from the IFS data, using only the data from the H band part of the spectrum. To this aim we transformed the two images in polar coordinates and on these images we performed a cross-correlation to find the relative rotation of one image with respect to the other, masking all but the region around the compact feature. We find a rotation of $1.31^{\circ}\pm 0.08^{\circ}$ in the clockwise direction that is in good agreement with the Keplerian rotation of around $1^{\circ}$ 
expected for an object at the separation of this feature, adopting the stellar mass from \citet{2021ApJ...908...46B}. 

To further characterize the environment around DR\,Tau we obtained the azimuthal brightness profiles at separations 0.3\as, 0.5\as 0.7\as and 0.9\as from the star with step of $2^{\circ}$ in PA. The results of this procedure are displayed in Figure~\ref{f:aziprof}. The plot obtained at a separation of 0.3\as is dominated by the peak due to the compact feature described above at a PA of $\sim$$20^{\circ}$ (indicated by a black arrow) while the enhanced flux due to the north-eastern spiral arm is also visible. At larger separations the azimuthal profiles are dominated by the peaks caused by the southern spiral arm. These peaks of emission are moving west at increasing separation from the star. The feature at a PA of $\sim$$110^{\circ}$ visible in the plot corresponding to a separation of 0.5\as might be an extension of the north-eastern spiral arm.

\subsection{Mass limits}
\label{s:masslimit}

\begin{figure}
\centering
\includegraphics[width=1.\columnwidth]{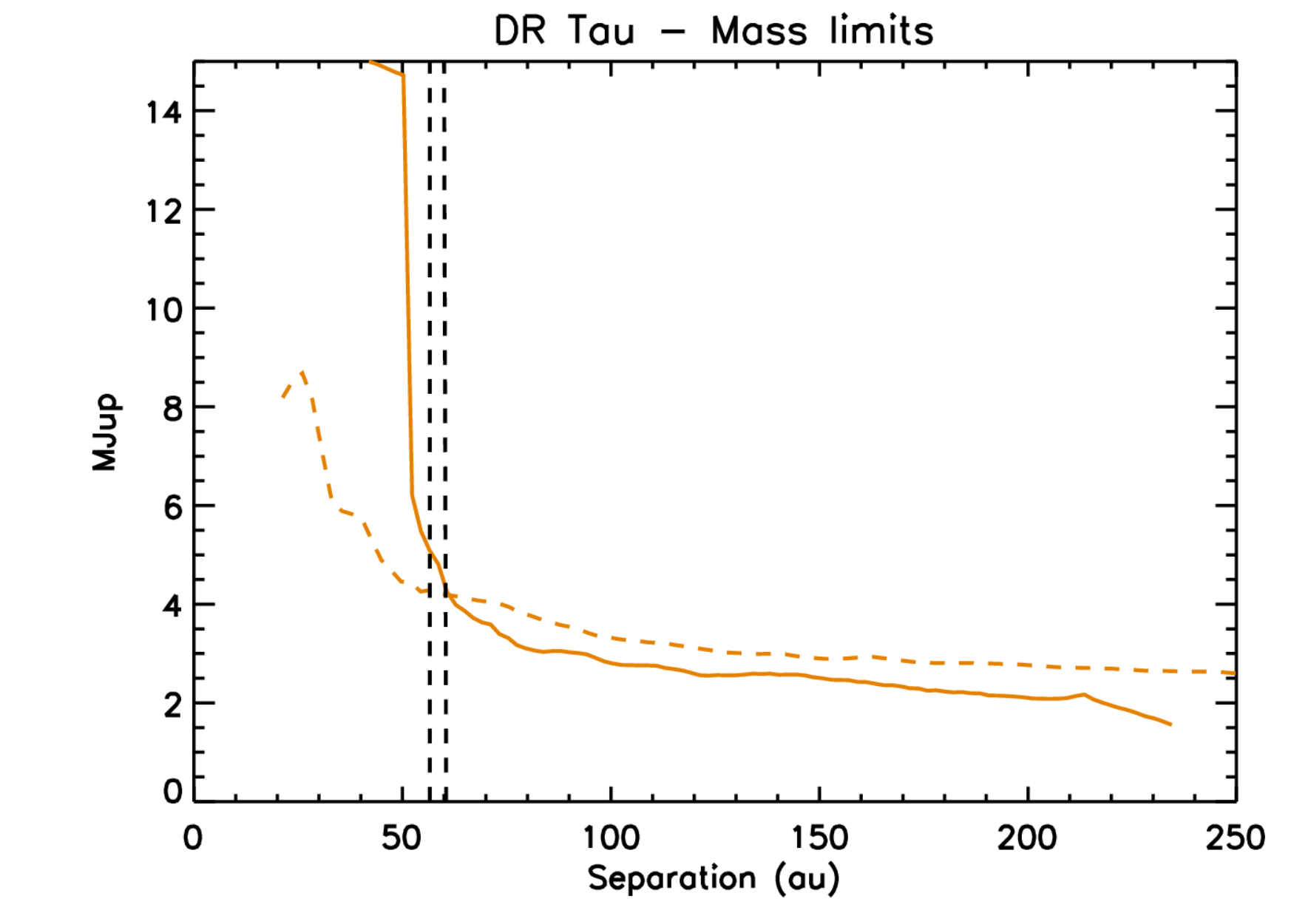}
\caption{Plot of the mass limits around DR\,Tau as obtained from the LMIRCam L' data (orange solid line) using the AMES-DUSTY models and assuming a system age of 3 Myr.
The dashed orange line represents the limits obtained using the total intensity IRDIS data assuming the same model and the same system age.
The two black dashed lines give the estimated separation of the compact feature.}
\label{f:masslimit}
\end{figure}

We calculated the contrast both for the IRDIS total intensity data and for the LMIRCam data applying the procedure devised in 
\citet{2015A&A...576A.121M}. The self-subtraction related to the high-contrast method was estimated injecting simulated planets at 
different separation in the original dataset and the contrast was then corrected accordingly. Finally, we corrected the results for the
small sample statistics following the method described by \citet{2014ApJ...792...97M}. We then transformed these contrast values in mass 
limits adopting the AMES-DUSTY models  \citep{2001ApJ...556..357A} and an age of 3~Myr. The results are shown in Figure~\ref{f:masslimit}
where we also show with two vertical dashed lines the estimated separation from the star of the
compact feature as obtained in Section~\ref{s:morpho}. In both cases the mass limits are of the order of few \MJup at large separations 
from the star while as expected the limits are
much lower for the SPHERE data at small separations from the star. We note in any case that, especially for the SPHERE case, the presence
of the disk can influence the results that should be regarded as rough estimates of the limits. 


\section{Discussion}
\label{s:dis}

\subsection{Origin of the southern spiral arm}
\label{s:orispir}

A possible explanation for the shape of the spiral arms is that they are due to a companion embedded in the disk. We will discuss this possibility for what concerns the north-eastern spiral arm in Section~\ref{natps}. For what concerns the southern spiral arm, we note that the pitch angle of this arm is large ($\sim$$26^{\circ}$) and increases with separation for any possible geometry. A prediction of the models from  \citet{2018ApJ...859..119B} is that the pitch angle of planet-driven spiral arms decreases with increasing separation from the planet. This would indicate that if the southern spiral arm is generated by a perturbing object, this should be located further than the spiral arm itself. This would imply a separation of the order of more than 200~au for
this putative companion. From the mass limit displayed in Figure~\ref{f:masslimit} we can exclude the presence of objects with mass larger than few \MJup. Companions with lower masses
could however be responsible for launching this spiral. In any case we note that for such low masses the theory of density wave caused by a gravitational perturber in the disk 
\citep{2002ApJ...569..997R} should be valid and we should be able to apply the formula defined by \citet{2012ApJ...748L..22M} to fit the spiral. However, to be able to fit 
a spiral with such high pitch angle we need to use a very high and unreliable value for the disk aspect ratio. Furthermore, the clear
bending toward north of this spiral described in Section~\ref{s:spiralarm} makes very improbable that it can be caused by a single
bound object in the disk but it would require the presence of multiple perturbing objects. These results make improbable that the southern spiral arm is generated by the perturbation of a planet.

The large value of the pitch angle also does not agree with predictions for spiral arms caused by shadows \citep{2016ApJ...823L...8M}.

As an alternative scenario, spiral arms can be caused by gravitational instability \citep[GI - see e.g., ][]{2015ApJ...812L..32D}. Through detailed simulations, this study concluded that this mechanism requires that the spirals are relatively compact, on scales less than 100~au, that the disk is massive with a mass ratio with the star $q$ larger than 0.1 
and that the accretion rate is of the order of $10^{-6}$~\MSun~yr$^{-1}$. These properties do not match the case of the southern arm of DR~Tau as it extends out to $\sim$220~au from the star, the stellar mass accretion rate is lower than required \citep[see discussion in][]{Antoniucci2017}, and moreover the disk mass is much less than needed. Indeed, \citet{2019AJ....157..144B} estimated a mass of the dust in the disk of $\log{M_{\rm dust}}=-3.53$ in solar units. If we adopt the usual dust to gas conversion of $M_{\rm gas}/M_{\rm dust}$=100, the disk mass is 
$M_{\rm Disk}=0.03$~\MSun, that is $q\sim0.03$. The same value is given by \citet{2019ApJ...872..158A}. This is an order of magnitude lower than required for the GI scenario. 
Furthermore, we could consider the value of the Toomre parameter (Q) that is commonly used to quantify the gravitational disk
stability \citep{1964ApJ...139.1217T}. Values of Q above 1.7 are indicative of a stable disk both for linear and second-order
perturbations \citep[see e.g.,][]{2014prpl.conf..643H,2021arXiv210906433S}. A detailed calculation of this parameter
is outside the scope of the present work but we can estimate its value by dividing the disk aspect ratio (H/R) by the value of q that we 
have defined above. A common value of 0.1 for the aspect ratio would imply a value larger than 3 for Q. This put the DR\,Tau disk into 
the stability regime. Even if this cannot be considered as a definitive result because of the large uncertainties in this analysis, it 
is a further indication favouring a stable disk. 
Finally, simulations show that the number of spiral arms generated in the context of the GI depends from $\sim 1/q$ \citep{2009MNRAS.393.1157C,2015ApJ...812L..32D,2019ApJ...871..228H}. Considering 
the value of q$\sim$0.03 determined above, we should expect for DR\,Tau around 30 spiral arms. This is a disk morphology completely different from what our data are showing.
For these reasons we then consider this scenario improbable.

Spirals in a disk may be generated by perturbation by a massive object that had a near encounter ($<$1000~au) in the recent past \citep[see e.g., ][]{2014MNRAS.441.2094R}. An analysis of the Gaia eDR3 proper motion of the other components of the small group in which DR\,Tau is located allowed us to find the stars that had the nearest approaches in the recent past. We found that there was a passage of DQ\,Tau at a projected separation of $26\pm2$\as (corresponding to 5.1$\pm$0.3~kau) about 0.231$\pm$0.007~Myrs ago. 
While we can provide only an estimate due to the uncertainties on its proper motion, the close binary V\,1001\,Tau might have had an 
even closer encounter with DR\,Tau though longer ago, passing at a projected separation of only 6\as (that is, $\sim$1.1~kau) about 
0.850~Myrs ago. No other members of the group with entries in Gaia eDR3 passed at smaller distance from DR\,Tau. 
In this analysis we did not consider the component along the line of sight of the velocities as the error bars on the parallax are 
of the order of 30~$\mu$as that, at the distance of the system, corresponds to $\sim$1.2~pc. This is much larger than the minimum 
separations we calculated. This makes the use of the velocity along the line of sight unreliable in this context.
While the two passages described above
were possibly close enough to trigger a spiral pattern in the DR\,Tau disk, it is unlikely that this pattern could be visible for such a 
long time after the passage because of disk viscosity in the inner region, and because the material extracted by the passage would have 
spread over more than 10~kau after $10^5$~yr \citep[see][]{2014MNRAS.441.2094R}. These results, in association with the lack of any 
possible close passing object in IRDIS FoV even much fainter than detectable by Gaia eDR3 (itself as deep as 20 $M_{\rm Jup}$ at 
separation $>3$~arcsec), make this explanation for the shape of the spiral pattern unlikely.

\begin{figure}
\centering
\includegraphics[width=1.\columnwidth]{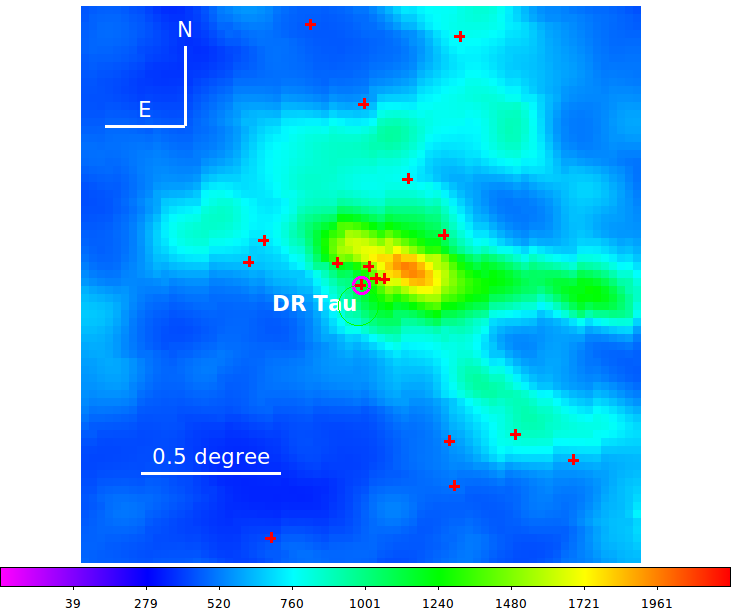}
\caption{Map of thermal dust emission in the direction of DR Tau obtained by Planck including the Lynds 1558 cloud. The colour code represents intensity. Red crosses are position of stars projected close to DR\,Tau, that is marked with a red circle; however only those stars closer to DR\,Tau are likely members of the small association around DR\,Tau}
\label{f:planck}
\end{figure}

A last scenario that we considered requires the presence of infalling material from late encounters of the star with low mass cloudlets as it moves through the molecular cloud in which it formed as recently proposed by models by e.g. \citet{2019A&A...628A..20D} and \citet{2020A&A...633A...3K}. According to what was found by the models cited above, the infalling material might have a very different angular momentum with respect to the star and to the material surrounding it. For this reason this material would not fall directly onto the star but would orbit around it leading to the formation of a new disk or to the replenishment of the existing one with a clear misalignment of the new structure with respect to the structures formed during the star formation. The formation of spirals by the infalling material is a natural consequence of this process. The clear changes in surface brightness for the spirals of DR\,Tau could be due to the fact that the spiral itself is distributed on different planes and with different inclination with respect to the star. This configuration favours visibility in scattered light and helps explaining why the corresponding emission is not detected in the ALMA continuum data \citep{2019ApJ...882...49L}, though we notice that material as far from the star as the southern spiral was detected by ALMA in molecular lines \citep{2021ApJ...908...46B}. We note moreover that according to the results of the simulations cited previously, not all the material of the cloudlets is captured by the star. This remaining material performs a flyby with curved trajectory around the star that results in an arc-like structure resembling the structure that we actually identified in the polarized data of DR\,Tau  and indicated by a solid red line in Figure~\ref{f:spiral}. \par
The native environment of DR\,Tau makes this hypothesis further possible. Indeed, the star is located in the close vicinity of Lynds\,1558 \citep{1962ApJS....7....1L} that likely is the parent cloud of the small association including the star \citep{2009ApJ...694.1423L}. Unfortunately this region is slightly out of the high resolution maps obtained with Herschel and described by \citet{2020A&A...638A..85R}. These maps however show a filamentary structure in the direction from north-east to south-west in an adjacent region. At lower resolution, Figure~\ref{f:planck} shows the environment of DR Tau in the thermal dust emission map from the 2018 version of the Planck Legacy Archive (\url{http://pla.esac.esa.int/pla/#maps}). From this map, it appears that DR\,Tau is located at the south-eastern edge of Lynds\,1558, in agreement with the idea that the formation of the cloud and of the stars could be triggered by the expansion of the Orion super-bubble \citep{2009ApJ...694.1423L}. Within this environment, late accretion of material to DR\,Tau looks well possible. More likely, this material should come from north-west where the densest region of Lynds\,1558 is located, but this could be properly assessed only if the relative motions were known. At the moment this is not possible as the spatial velocities UVW
for Lynds\,1558 are unknown \citep[e.g.,][]{2019A&A...630A.137G}.\par
All these characteristics of the DR\,Tau disk tend to favour an interpretation of infalling material responsible for creating at least the southern spiral. The number of spiral disks for which the late infalling of material has been proposed to explain their shape is at the moment small. We can cite e.g. AB\,Aur, HD\,100546  \citep{2019A&A...628A..20D}, HL\,Tau \citep{2019ApJ...880...69Y}, and SU\,Aur \citep{2021ApJ...908L..25G}. All these stars have a spiral pattern comparable to that of DR\,Tau but, on the other hand, they also have a mass larger than 2~\MSun. According to \citet{2019A&A...628A..20D} the presence of such structures around stars with lower masses should be less probable. DR\,Tau should then be one of the lower mass stars around which these structures are actually present. 

\subsection{Nature of the compact feature and of the north-eastern spiral arm}
\label{natps}

\begin{figure}
\centering
\includegraphics[width=1.\columnwidth]{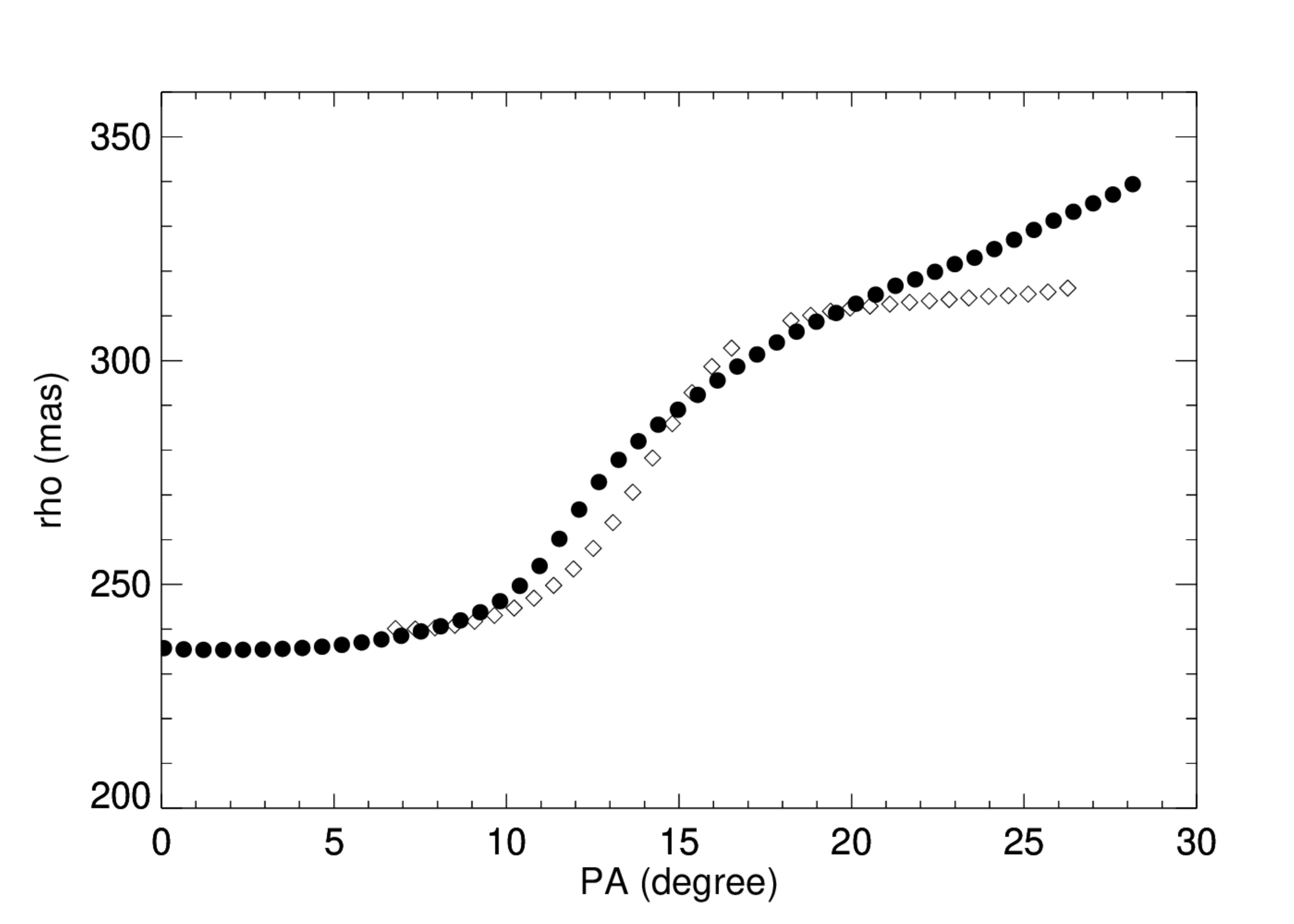}
\caption{Plot of the polar coordinates of the eastern spiral arm around the position of the compact feature. The full circles represents the positions obtained with IFS data while the empty diamonds represent the positions obtained with the IRDIS polarized data. Error bars are too small with respect to the scale of the plot and are not visible.}
\label{f:polarspiral}
\end{figure}

\begin{figure*}[htp]
\centering
\includegraphics[width=0.9\textwidth]{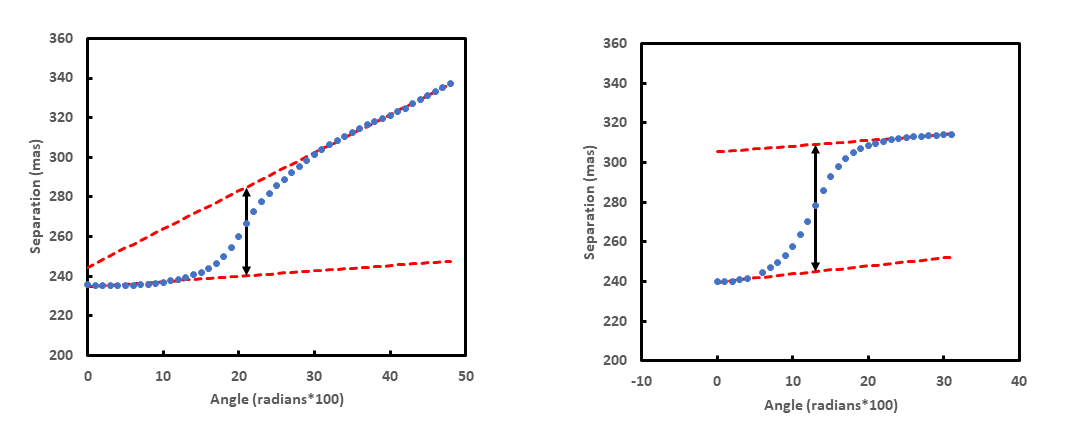}
\caption{Linear fit (dashed red lines) of the inner and of the outer part of the westward spiral (blues filled circles). The black arrow indicates the offset between the two linear fits. The procedure is illustrated both for the IFS H spectral band data (left panel) and for the IRDIS polarized data (right panel).}
\label{f:spine}
\end{figure*}

As seen above, a compact structure is visible north of the star at both the SPHERE epochs and both in polarized light and in total intensity. 
This fact makes it very probable that we are looking at a real feature of the disk and not simply at a structure created by the differential imaging data reduction methods. In the 
polarimetric data, this feature seems connected to the north-eastern spiral arm. While all the results hint toward stellar light reflected by dust, the nature of this bright compact structure is not clear. One intriguing possibility, that is reinforced by the Keplerian motion tentatively detected for this feature by the analysis described in Section~\ref{s:morpho}, is that it is caused 
by the presence of a planet in formation and still embedded in its dust envelope. 

The characteristics of the feature described in Section~\ref{s:morpho} resemble those of one of the two features detected by \citet{2020A&A...637L...5B} in the AB\,Aur system whose disk is also characterized by the presence of spiral arms. Indeed, in both cases the feature appears to be elongated and, due to their detection both in polarized and total intensity light, it is improbable that we are looking at the emission from the photosphere of a planet, at least at the wavelengths imaged by SPHERE. Furthermore, like for the case of AB\,Aur, this feature is within one of the two spiral arms detected in the disk itself. The most probable explanation for the nature of the feature in AB\,Aur was that it was caused by a forming planet. However, we note that in that case the authors were able to fit the spirals shape with the formula from \citet{2012ApJ...748L..22M} while we found that this was not possible for the case of DR\,Tau. For this reason the similarity between the two cases should be taken with caution.

In any case to strengthen the hypothesis of the presence of a companion we can consider the shape of the spiral around the position of the proposed companion. To this aim we have deprojected the image to the plane of the disk using the ALMA values, we have transformed it into polar coordinates and for each position angle (with a step of 0.01 radians) we have calculated the position of the spiral considering the peak of a Gaussian fit in the radial direction and assuming as error the FWHM of the gaussian fit. The same procedure has been applied both to the IFS Y-H data and to the IRDIS polarized data. In this way it was possible to compare the shape of the spiral in the two epochs after shifting the IFS data by $1.31^{\circ}$ to account for the shift between the two epochs measured in Section~\ref{s:morpho}. Results at different epochs that are displayed in Figure~\ref{f:polarspiral} are very similar each other and, more importantly, they both exhibit the S shape that is characteristic of the presence of a companion according to the results of hydrodynamic simulations \citep[see e.g. ][]{2015ApJ...813...88Z,2018ApJ...859..119B}.

If the presence of a companion is the correct explanation for this structure it would be important to estimate its mass. This is however a very difficult task given that, as explained before, we are not directly seeing the planetary photosphere. Since radiation at longer wavelengths is less absorbed or scattered and a small mass planet is likely very cool, a constraint could be obtained from the mass limit obtained from the observation in L' spectral band with LMIRCam described in Section~\ref{s:lmircam} and  shown in Figure~\ref{f:masslimit} with 
a solid lines. While these results should be taken with some care due to the bad weather conditions in which the data were taken, they can however be useful to give some indication on the mass of the putative object. From the plot it is apparent that the mass of the companion should be less than a few \MJup otherwise it should be visible also in the LMIRCam data.

A further indication that the mass of the companion should be small is that we are not able to detect any gap in the disk in our 
images of the DR\,Tau system. The presence of such gap would be requested for a companion with a mass larger than few \MJup embedded in 
the disk according e.g., to the formula provided in \citet{2016PASJ...68...43K}. However, we note that the ALMA mm emission outer 
radius is inside of the current companion position. Then, the companion may rather reside on the outside of the dust disk, truncating it. 
In this case we would not expect to see any gap-like structures. From the ALMA continuum data it is however not clear if the gas disk 
extends further out than the bulk of the mm-sized grains.

We can also use the shape of the north-eastern spiral to confirm this results. Indeed, the radius of the Hill sphere of the proposed 
companion should be roughly half of the offset in radial position of the leading and trailing part of the spiral arm at its position
\citep{2020A&A...642A.187S}. To take into account the pitch angle, we computed linear fits of the leading and trailing portion of the 
spiral arm and measured the shift between the two fitting lines at the proposed companion position. We applied this procedure both to IFS
H-band data and IRDIS polarized data, as illustrated in Figure~\ref{f:spine} where we represent with blue filled circles the spiral 
points and with red dashed lines the results of the linear fit procedure. The black arrows represent the measured offset between the 
inner and the outer part of the spiral. From this procedure we obtain for the Hill radius a value of 22~mas in the case of the IFS data 
and of 32~mas in the case of the IRDIS polarized data. We can then assume for the Hill radius of the proposed companion a value of 
$27\pm5$~mas that corresponds to a mass of the embedded companion of $2.5^{+1.7}_{-1.1}$~\MJup if we assume for the stellar mass the 
value obtained by \citet{2021ApJ...908...46B}. 
In any case the previous discussion cannot be considered as a precise measure of the Hill radius and, as a consequence, of the
mass of the companion. Indeed, in the case of a small companion not opening a gap, the launching position for a spiral density wave
could not be at the separation of the Hill radius but at a separation corresponding to 2/3 of the disk scale height \citep{2015ApJ...815L..21F}
which is larger than the Hill radius of low mass planets. For this reason
the value obtained using the previous considerations has to be considered as a rough estimation of the mass. In any case it can
give important indications about the order of magnitude of the companion mass and it is important to confirm that it should be of the
order of few \MJup.

Finally, we note that if the embedded companion were more massive than a few Jupiter mass, we should expect to have a symmetric twofold spiral pattern \citep{2018ApJ...859..119B} that we actually do not see. We conclude that the properties of the compact feature north of the star are compatible with the presence of an embedded planetary mass companion, with a mass of a few Jupiter masses and that this planet could be responsible for the shape of the nort-eastern spiral. 

While the evidences presented above seem to point toward a spiral caused by the presence of a planet, we cannot in any case
exclude any other possible cause for the formation of the north-eastern spiral. In particular if, as we discussed in Section~\ref{s:orispir},
the southern spiral is probably caused by late infall of material in the system, we cannot exclude that also the north-eastern spiral
has been affected at least partially by this process. We could also consider a scenario in which both planetary formation and late
infall of material are involved in the definition of the shape of this spiral. 
In this context we could explore a scenario that considers the formation and growth of planets in disks with late accretion of material, e.g. in the context of planetesimals \citep{1996Icar..124...62P}, pebbles \citep[see e.g., ][]{2010MNRAS.404..475J,2019A&A...623A..88B}, or hybrid \citep[see e.g., ][]{2018NatAs...2..873A} accretion models . While seeds for planet formation are likely needed anyway, they may be much smaller than the final planet and may form closer to the star where timescales for seed core formation is fast enough. During the late infall episodes, the gas and pebble flux may grow considerably, though temporarily, in the outer regions favouring core growth and then gas runaway accretion. Of course these scenario are complicated by further migration due to interaction between the planet and the disk and by planet-planet scattering, but in favourable circumstances they might possibly lead to the formation of giant planets at large separation. Due to the large uncertainties, such scenarios can only be supported by snapshot observation of the process while in progress.


\section{Conclusions}
\label{s:conclusion}

We presented new observations of the DR\,Tau system with both SPHERE at VLT and with LBTI/LMIRCam at LBT. 
The main results obtained from the analysis of these data are summarized here:
\begin{itemize}
\item The SPHERE polarized data in the H band allowed us to detect a previously undetected system of two spirals around this star, north-east and south of the star, respectively. The same spirals are also visible in the total intensity SPHERE data even if with much lower signal-to-noise. 
\item According to our analysis, the most probable origin of the southern spiral is late infalling of material from cloudlets present in the formation environment of the star. The presence of a clear arc-like structure just north to the star is a further confirmation of this possibility as also this structure is foreseen in this scenario. We moreover excluded with a good degree of certainty other possible scenarios to explain this spiral arm.
\item On the other hand, the possibility that the north-eastern spiral is caused by the presence of an embedded companion cannot be excluded. Indeed, a candidate for this explanation is present just north to the star at separation of $\sim$0.3\as where a compact bright structure is clearly visible in the SPHERE data both in polarized and non-polarized light. This result and the blue SPHERE spectrum of this feature make it probable that we are not seeing the photosphere of a planet but dust illuminated by the light from the central star. 
\item The nature of this compact structure is unclear but it strongly resembles a feature identified around AB\,Aur, which also hosts an extended spiral system. In that case this structure was
interpreted as a planet in formation still embedded in its dust envelope. 
\item The planetary nature of this structure is further supported by the S shape of the north-eastern spiral itself in the region just around its position. This shape is in good agreement with that expected around an embedded companion resulting from hydrodynamic simulations. 
\item The possibility that the structure of the north-eastern spiral and the compact feature is influenced by late infall as happens for the southern spiral cannot be excluded and we also considered the possibility the both processes are involved in this case.
\item The mass of this putative companion may be up to the order of few \MJup considering the upper limits provided by the L'
observations with LBTI/LMIRCam according to which a companion with a larger mass should be detected. This result is strengthened by
the absence of a clear gap in the disk that would be opened by a larger mass companion and by the asymmetry of the spiral pattern that
should be symmetric in case of a larger mass companion. Finally, an estimation of the Hill radius based on the offset of the spiral 
around the companion position confirms that the order of magnitude of the companion mass is in the range of few \MJup or less.
\end{itemize}

While the high spatial resolution offered by SPHERE is crucial in deriving the geometry of the compact feature, observations at longer wavelengths are needed to disentangle light of a possible companion from the dust in which it is embedded. A first attempt in this direction was done with our observations of DR\,Tau in the L' band. Unfortunately, mainly due to bad weather conditions, these observations were not conclusive resulting in non-detection of both companion and disk. Future similar observations both with LMIRCam or with similar instruments will be very useful to give a conclusive solution to this problem. Finally, we might expect that if really present, the planet should be accreting and possibly detectable through H emission lines \citep[e.g., ][]{2018ApJ...863L...8W,2019NatAs...3..749H}. While they can be possibly strongly absorbed by the circumplanetary material in the visible, they might be detectable in the near IR.


We note that if the results of this work were confirmed, DR\,Tau would be of paramount importance. This system is much younger than PDS\,70, currently the benchmark for very young planets caught at very early phases of formation. Indeed in the case of PDS\,70, the mass of the disk of $\sim 0.003$~\MSun \citep{2018A&A...617A..44K} is an order of magnitude smaller than the total mass of the two planets \citep{2018A&A...617A..44K,2019A&A...632A..25M}, while in the case of DR\,Tau the mass of the disk of $\sim 0.03$~\MSun \citep{2019ApJ...872..158A} is an order of magnitude larger than the mass of the possible planet. DR\,Tau  is then observed in a much earlier phase, consistent with the age estimates for the stars of 6~Myr for PDS\,70 \citep{2018A&A...617L...2M} and $<3$~Myr for DR\,Tau (see Section~\ref{target}). Furthermore, together with AB\,Aur \citep{2020A&A...637L...5B} and perhaps HD\,100546 \citep{2013ApJ...766L...1Q, 2018A&A...619A.160S, 2019A&A...628A..20D}, DR\,Tau would be one of the few known objects for which ongoing planetary formation and infall of material on the disk might be present at the same time. This may be relevant to understand how massive planets can form at very large distances from the star. This is notoriously difficult to explain in scenarios where planets form within a given disk, where the total system mass is constant \citep[see e.g. the discussion in ][]{2019AJ....158...13N}. Given its very young age and distance from its star, which has a mass comparable to that of the Sun, the proposed companion of DR\,Tau would then be an important challenge for the current scenarios of planet formation. 


\begin{acknowledgements}
  The authors thank the referee for the constructive comments that helped to improve this work.

  This work has made use of the SPHERE Data Center, jointly operated by OSUG/IPAG (Grenoble), PYTHEAS/LAM/CeSAM (Marseille), OCA/Lagrange (Nice) and Observatoire de Paris/LESIA (Paris). \par
This work has made use of data from the European Space Agency (ESA) mission {\it Gaia} (\url{https://www.cosmos.esa.int/gaia}), processed by the {\it Gaia} Data Processing and Analysis Consortium (DPAC, \url{https://www.cosmos.esa.int/web/gaia/dpac/consortium}). Funding for the DPAC has been provided by national institutions, in particular the institutions participating in the {\it Gaia} Multilateral Agreement. 

This work was supported by the PRIN-INAF 2019 Planetary Systems At Early Ages (PLATEA). 

This research has made use of the SIMBAD database, operated at CDS,
Strasbourg, France. 

D.M., R.G., S.D., A.Z. acknowledge support from the ``Progetti Premiali'' funding scheme of the Italian Ministry of Education, University, and Research. 
A.Z. acknowledges support from the CONICYT + PAI/Convocatoria nacional subvenci\'on a la instalaci\'on en la academia, convocatoria 2017 + Folio PAI77170087. 
A.M. acknowledges the support of the DFG priority program SPP 1992 "Exploring the Diversity of Extrasolar Planets" (MU 4172/1-1). C.\,P. acknowledge financial support from Fondecyt (grant 3190691) and financial support from the ICM (Iniciativa Cient\'ifica Milenio) via the N\'ucleo Milenio de Formaci\'on Planetaria grant, from the Universidad de Valpara\'iso. T.H. acknowledges support from the European Research Council under the Horizon 2020 Framework Program via the ERC Advanced Grant Origins 83 24 28.

SPHERE is an instrument designed and built by a consortium consisting of IPAG (Grenoble, France), MPIA (Heidelberg, Germany), LAM (Marseille, France), LESIA (Paris, France), Laboratoire Lagrange (Nice, France), INAF-Osservatorio di Padova (Italy), Observatoire de Gen\`eve (Switzerland), ETH Zurich (Switzerland), NOVA (Netherlands), ONERA (France) and ASTRON (Netherlands), in collaboration with ESO. SPHERE was funded by ESO, with additional contributions from CNRS (France), MPIA (Germany), INAF (Italy), FINES (Switzerland) and NOVA (Netherlands). SPHERE also received funding from the European Commission Sixth and Seventh Framework Programmes as part of the Optical Infrared Coordination Network for Astronomy (OPTICON) under grant number RII3-Ct-2004-001566 for FP6 (2004-2008), grant number 226604 for FP7 (2009-2012) and grant number 312430 for FP7 (2013-2016). 

The LBT is an international collaboration among institutions in the United States, Italy and Germany. LBT Corporation partners are: The University of Arizona on behalf of the Arizona Board of Regents; Istituto Nazionale di Astrofisica, Italy; LBT Beteiligungsgesellschaft, Germany, representing the Max-Planck Society, The Leibniz Institute for Astrophysics Potsdam, and Heidelberg University; The Ohio State University, representing OSU, University of Notre Dame, University of Minnesota and University of Virginia. 

The results reported herein benefited from collaborations and/or information exchange within NASA’s Nexus for Exoplanet System Science (NExSS) research coordination network sponsored by NASA’s Science Mission Directorate. K.W. acknowledges support from NASA through the NASA Hubble Fellowship grant HST-HF2-51472.001-A awarded by the Space Telescope Science Institute, which is operated by the Association of Universities for Research in Astronomy, Incorporated, under NASA contract NAS5-26555.
\end{acknowledgements}

\bibliographystyle{aa}
\bibliography{drtau}

\end{document}